\documentclass[a4paper,11pt]{article}
\usepackage{pos}

\newcommand{\be}{\begin{equation}}
\newcommand{\ee}{\end{equation}}
\newcommand{\bea}{\begin{eqnarray}}
\newcommand{\eea}{\end{eqnarray}}

\title{
A novel method to evaluate real-time path integral for scalar $\phi^4$ theory
}
\ShortTitle{
A novel method to evaluate real-time path integral for scalar $\phi^4$ theory
}

\author*{Shinji Takeda}

\affiliation{
Institute for Theoretical Physics, Kanazawa University,\\
Kanazawa 920-1192, Japan
}

\emailAdd{takeda@hep.s.kanazawa-u.ac.jp}

\abstract{
We present a new scheme which numerically evaluates the real-time path integral for $\phi^4$ real scalar field theory in a lattice version of the closed-time formalism.
First step of the scheme is to rewrite the path integral in an explicitly convergent form by applying Cauchy's integral theorem to each scalar field. 
In the step an integration path for the scalar field is deformed on a complex plane such that the $\phi^4$ term becomes a damping factor in the path integral.
Secondly the integrations of the complexified scalar fields are discretized by the Gauss-Hermite quadrature and then the path integral turns out to be a multiple sum.
Finally in order to efficiently evaluate the summation we apply information compression technique using the singular value decomposition to the discretized path integral, then
a tensor network representation for the path integral is obtained after integrating the discretized fields.
As a demonstration, by using the resulting tensor network we numerically evaluate the time-correlator in 1+1 dimensional system.
For confirmation, we compare our result with the exact one at small spatial volume.
Furthermore, we show the correlator in relatively large volume using a coarse-graining scheme
and verify that the result is stable against changes of a truncation order for the coarse-graining scheme.
}

\FullConference{
 The 38th International Symposium on Lattice Field Theory, LATTICE2021
  26th-30th July, 2021
  Zoom/Gather@Massachusetts Institute of Technology
}

\begin{document}
\maketitle

\section{Introduction}
Study of real-time dynamics for quantum many-body system is one of the important topics of the modern physics.
Since real-time evolution of a system is governed by Hamiltonian, a natural choice for such a study would be the Hamiltonian formalism.
In fact, tensor networks using the Hamiltonian approach are applied in various models \cite{Pichler:2015yqa,Buyens:2016hhu,Banuls:2019ybd}.
Recently, quantum simulations of quantum field theory 
based on the Hamiltonian formalism also become popular \cite{Yamamoto:2020eqi,Gustafson:2021qbt}.

On the other hand, there is another approach based on the Lagrangian formalism, namely the path integral quantization.
To study real-time systems with the path integral approach,
the Schwinger-Keldysh formalism \cite{Schwinger:1960qe,Keldysh:1964ud} is established and by choosing a suitable initial density matrix one can treat
not only equilibrium but also non-equilibrium systems.
The approach, however, suffers from the notorious sign problem.
To overcome the problem, the complex Langevin method \cite{Berges:2005yt,Berges:2006xc}
and the Lefshetz thimble method \cite{Alexandru:2017lqr,Mou:2019tck,Mou:2019gyl} are proposed.

In this paper, we adopt the tensor network method based on the Lagrangian approach.
Although we actually presented a preliminary study using the Feynman prescription technique in the previous lattice conference \cite{Takeda:2019idb},
here we introduce a different attempt and
show some numerical results of the time-correlator.

\section{Closed-time formalism}
The path integral approach to real-time system is formulated by 
Schwinger and Keldysh \cite{Schwinger:1960qe,Keldysh:1964ud}
and its lattice version is discussed in \cite{Berges:2006xc,Alexandru:2017lqr,Hoshina:2020gdy,Kanwar:2021tkd}.
In the following we consider
1+1 dimensional lattice system, whose coordinate is denoted as $(t,n)$ with $t=0,1,2,\cdots,2N_{\rm T}+N_\beta-1$
and
$n=0,1,2,\cdots,N_{\rm S}-1$.
The periodic boundary condition is imposed for all directions.
The density matrix is set to the thermal one with the inverse temperature $\beta$.

For the real scalar field theory, the associated path integral on the lattice
is given by
\be
{\cal Z}
\equiv
\int [d\phi]
e^{iS_{\rm lat}[\phi]}
=
\int_{-\infty}^\infty
\prod_{t,n}
 d\phi_{t,n}
H^{[0]}(t;\phi_{t,n},\phi_{t+1,n})
H^{[1]}(t;\phi_{t,n},\phi_{t,n+1})
\label{eqn:path_integral}
\ee
where $H^{[\mu]}$ is a local transfer matrix for $\mu$-direction ($\mu=0$ and $1$),
\be
H^{[0]}(t;\phi,\phi^\prime)
\equiv
\exp
\left\{
-
\frac{\gamma}{2u_t}
(\phi-\phi^\prime)^2
-
\frac{u_t+u_{t-1}}{8\gamma}
V(\phi)
-
\frac{u_{t+1}+u_{t}}{8\gamma}
V(\phi^\prime)
\right\},
\ee
\be
H^{[1]}(t;\phi,\phi^\prime)
\equiv
\exp
\left\{
-
\frac{u_{t}+u_{t-1}}{4\gamma}
(\phi-\phi^\prime)^2
-
\frac{u_{t}+u_{t-1}}{8\gamma}
V(\phi)
-
\frac{u_{t}+u_{t-1}}{8\gamma}
V(\phi^\prime)
\right\}.
\ee
In the above equations, we have defined the anisotropic parameter
$\gamma=a/a_{\rm t}$ where
$a$ ($a_{\rm t}$) is a lattice spacing for the spatial (temporal) direction, and
the scalar potential is given by $V(\phi)=m^2\phi^2/2+\lambda\phi^4/4!$ where parameters $m$ and $\lambda$ are given in lattice units ($a=1$).
$u_t$ is a time-dependent phase factor that encodes the metric of the system and a shape of the closed-time path \cite{Hoshina:2020gdy}
\be
u_t
=
\left\{
\begin{array}{rlll}
 i  & \mbox{ for } & 0\le t < N_{\rm T} \\
 1  & \mbox{ for } & N_{\rm T}\le t < N_{\rm T}+N_\beta/2 \\
-i  & \mbox{ for } & N_{\rm T}+N_\beta/2\le t < 2N_{\rm T}+N_\beta/2 \\
 1  & \mbox{ for } & 2N_{\rm T}+N_\beta/2\le t < 2N_{\rm T}+N_\beta. \\
\end{array}
\right.
\ee
The physical real-time extent $T$ and the inverse temperature $\beta$ in lattice units
are given by the lattice parameters,
$
T=N_{\rm T}/\gamma
$
and
$
\beta=N_\beta/\gamma.
$

\section{Tensor network representation}
\begin{figure}[t!]
\begin{center}
\begin{tabular}{c}
\includegraphics[width=9cm,pagebox=cropbox,clip]{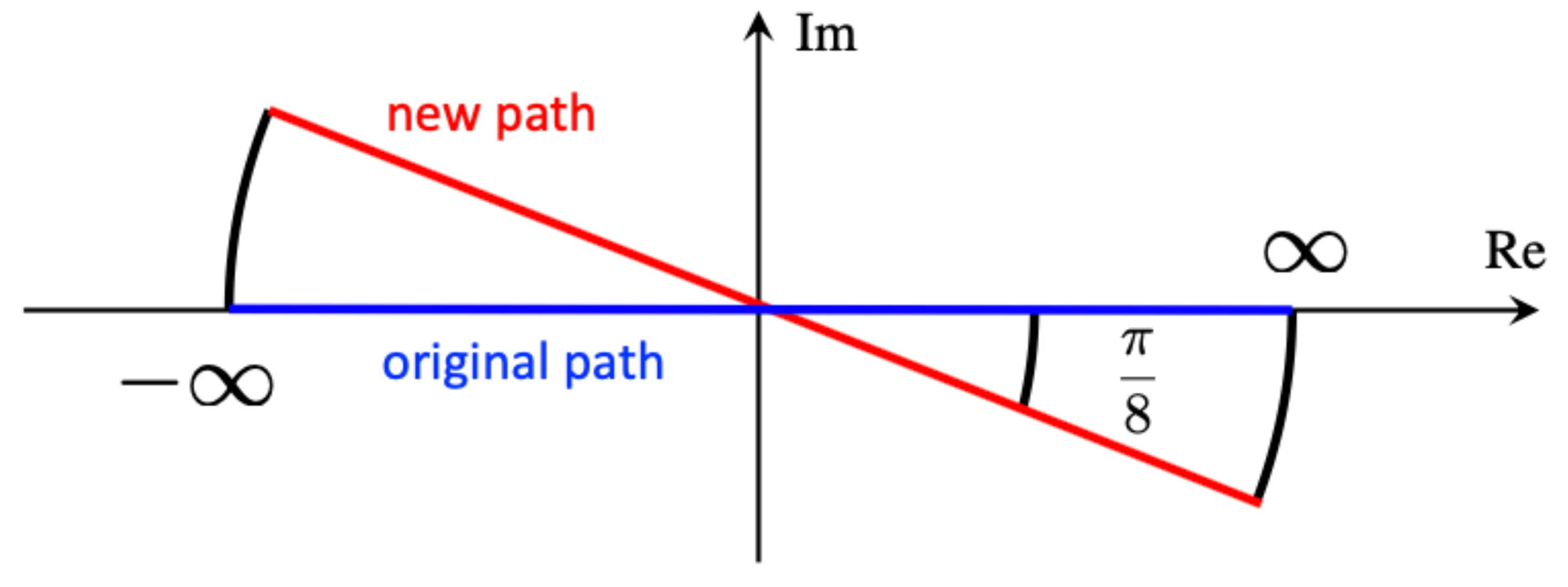}
\end{tabular}
\end{center}
\vspace{-5mm}
\caption{
Integration path in complex plane for a scalar field.
The blue horizontal line on the real axis is the original integration path. The red line tilted at $-\pi/8$ is a deformed new path.
}
\label{fig:complex_plane_scalar_field}
\end{figure}

The path integral in eq.(\ref{eqn:path_integral}) is given by multiple oscillating integrals and
at first glance, it is not clear that ${\cal Z}$ is convergent or not,
therefore, first of all we shall rewrite it with an explicitly convergent expression.
For that purpose, first
we consider a complex plane (see Fig.~\ref{fig:complex_plane_scalar_field}) for a scalar field $\phi$ that resides in the Minkowski region ($1\le t \le N_{\rm T}-1$).
Following Cauchy's integral theorem, we deform the original path to a new one tilted at $-\pi/8$ without changing a value of the integral,
namely, 
the associated part of the scalar field in the path integral, where the quartic potential term is explicitly written, is given by
\be
\int_{-\infty}^\infty d\phi
f(\phi)
e^{-i\lambda\phi^4/4!}
=
\int_{\mbox{new path}} d\tilde\phi
f(\tilde\phi)
e^{-i\lambda\tilde\phi^4/4!}
=
\int_{-\infty}^\infty d\xi
\frac{d\tilde\phi(\xi)}{d\xi}
f(\tilde\phi(\xi))
e^{-\lambda\xi^4/4!},
\ee
where a complex variable on the new path \footnote{
By modifying the new path, which is a straight line, one may improve an efficiency.
} is parameterized as $\tilde\phi(\xi)=e^{-i\pi/8}\xi$ with $\xi\in\mathbb{R}$.
Note that the last expression is clearly convergent for a given proper function $f$.
By applying such a deformation to every scalar field,
we can see that the path integral turns out to be explicitly convergent
\be
{\cal Z}
=
\int_{-\infty}^\infty \prod_{t,n} d\phi_{t,n}
e^{iS_{\rm lat}[\phi]}
=
\int_{-\infty}^\infty \prod_{t,n} d\xi_{t,n}
J(\xi)
e^{iS_{\rm lat}[\tilde\phi(\xi)]},
\ee
where $J$ is a total Jacobian and a phase factor in $\tilde\phi$ should be properly chosen depending on the time coordinate $t$.

We have seen that ${\cal Z}$ is well-defined and convergent, but its integrand is still complex value, so the Monte Carlo method cannot be applied.
In order to carry out the integration of $\xi$, we employ the Gauss-Hermite quadrature \cite{Kadoh:2018hqq}, i.e. for a proper function $g$,
an integral is replaced by a summation
\be
\int_{-\infty}^\infty
d\xi g(\xi)
\approx
\sum_{a=1}^N w_a e^{\xi_a^2}g(\xi_{a})
\ee
where $w_a$ are weight given by a known function, $\xi_a$ are zeros of $N$-th Hermite polynomial,
and 
a truncation order $N$ is a single parameter to control the degree of the approximation.
By applying such a quadrature rule to all $\xi$s,
the path integral now becomes a multiple sum
\be
{\cal Z}
=
\int_{-\infty}^\infty \prod_{t,n} d\xi_{t,n}
J(\xi)
e^{iS_{\rm lat}[\tilde\phi(\xi)]}
\approx
\sum_{\{a\}}
w_{a_{t,n}}
\exp\left(\xi_{a_{t,n}}^2\right)
J(\xi_a)
e^{iS_{\rm lat}[\tilde\phi(\xi_a)]}.
\ee

\begin{figure}[t!]
\begin{center}
\begin{tabular}{cc}
\includegraphics[width=7cm,pagebox=cropbox,clip]{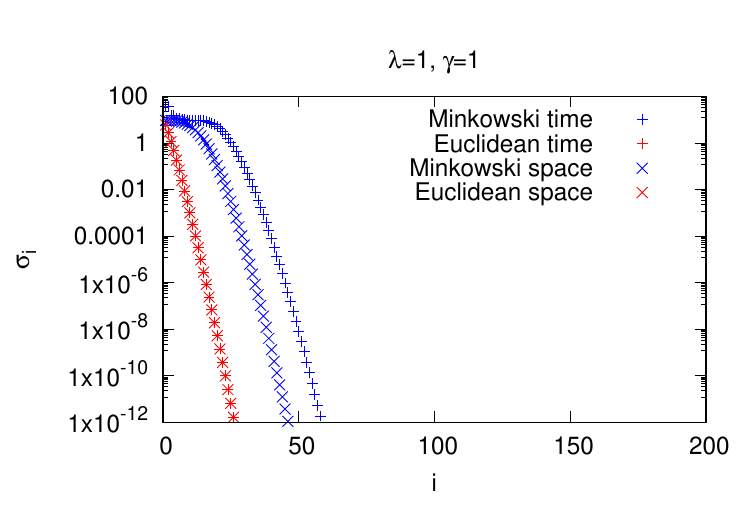}
&
\includegraphics[width=7cm,pagebox=cropbox,clip]{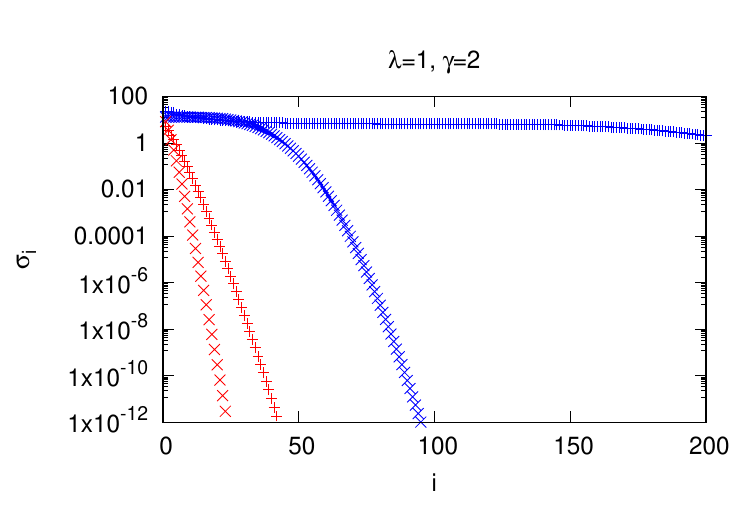}
\\
\includegraphics[width=7cm,pagebox=cropbox,clip]{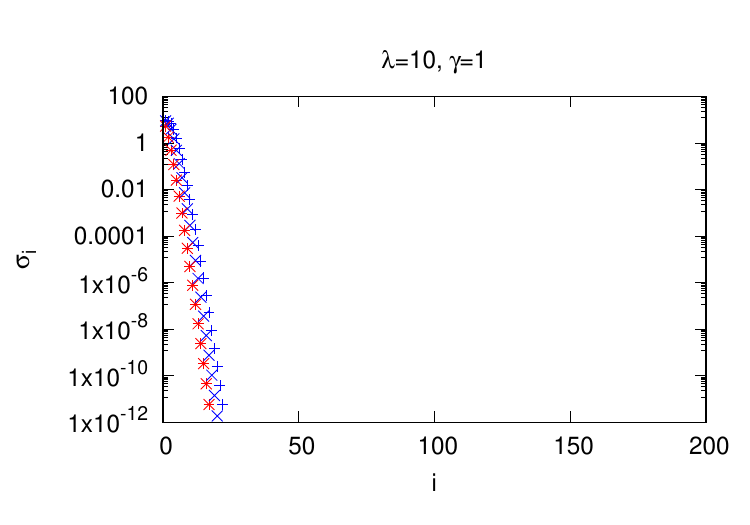}
&
\includegraphics[width=7cm,pagebox=cropbox,clip]{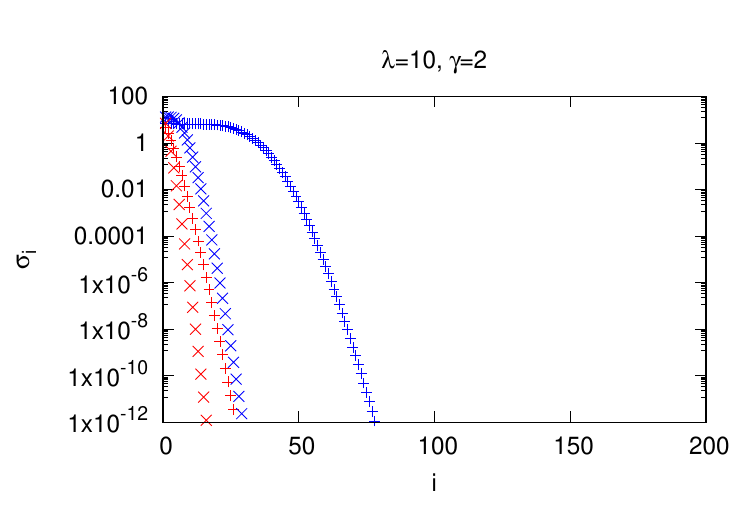}
\end{tabular}
\end{center}
\vspace{-5mm}
\caption{
Distribution of the singular values of $A^{[\mu]}(t)$ in Minkowski 
($1\le t\le N_{\rm T}-1$) 
and Euclidean 
($N_{\rm T}+1\le t\le N_{\rm T}+N_\beta/2-1$) 
region for time ($\mu=0$) and space ($\mu=1$) direction.
The mass parameter is set to $m=1$.
}
\label{fig:SVD}
\end{figure}

The discretization also allows us to define a matrix from $H^{[\mu]}$
\be
A^{[\mu]}_{ab}(t)
\equiv
H^{[\mu]}(t;\tilde\phi(\xi_{a}),\tilde\phi(\xi_{b}))
\left(w_{a} e^{\xi_{a}^2}\right)^{\frac{1}{4}}
\left(w_{b} e^{\xi_{b}^2}\right)^{\frac{1}{4}}.
\label{eqn:matrix_A}
\ee
Thanks to this, instead of dealing with continuous objects,
we can now use the linear algebra techniques.
The singular values of the matrix $A$ are shown in Fig.~\ref{fig:SVD}.
Basically the distribution of the singular values is wider (i.e. the correlation is stronger) in the Minkowski region 
than that of the Euclidean case. 
In the strong coupling case ($\lambda=10$),
the distribution is narrow, that is, the information is well compressed
and this is a nice feature from the computational point of view,
while in the weak coupling ($\lambda=1$) the distribution is wide hence a large number of modes are needed to maintain a precision.
For large anisotropic parameter ($\gamma=2$), the distribution tends to be wide compared with the isotropic case ($\gamma=1$),
thus the former case demands more computational cost than the latter.

\begin{figure}[t!]
\begin{center}
\begin{tabular}{c}
\includegraphics[width=14cm,pagebox=cropbox,clip]{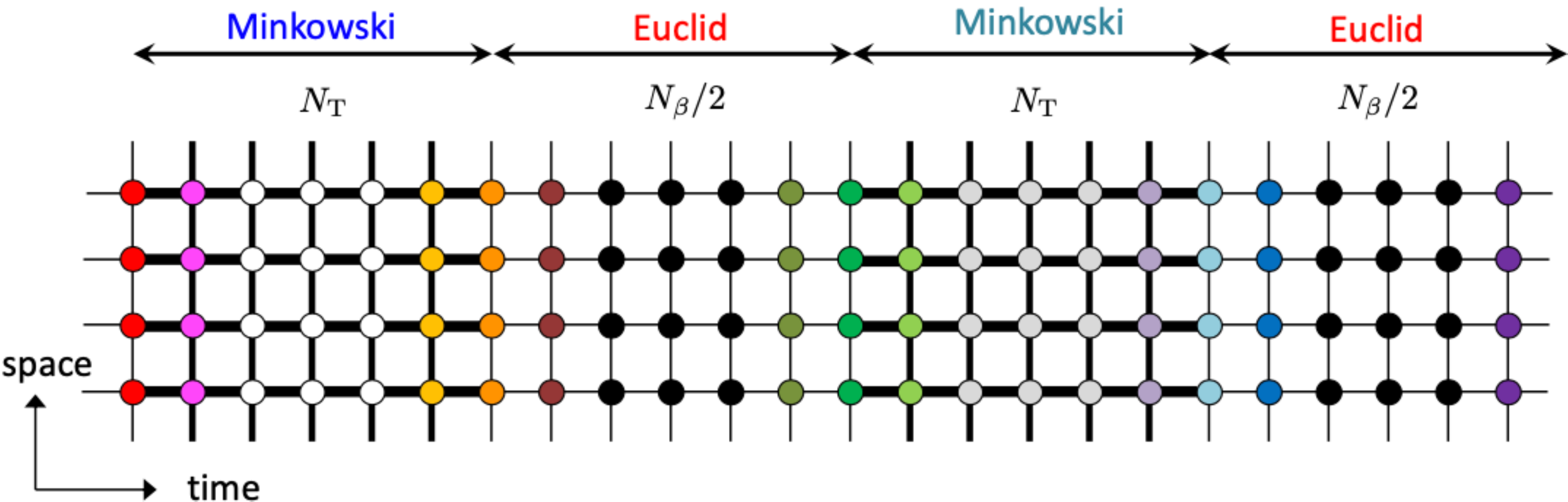}
\end{tabular}
\end{center}
\caption{
The tensor network for the system ($N_{\rm T}=N_\beta/2=6$, $N_{\rm S}=4$).
There are 16 kinds of tensors.
The strength of the correlation is represented by the width of the line.
}
\label{fig:tensor_network}
\end{figure}

As a final step, let us see how to make a tensor.
First, we apply the singular value decomposition to the matrix $A$
\be
A^{[\mu]}_{ab}(t)
=
\sum_{j=1}^{N}
U_{aj}^{[\mu]}(t)
\sigma_j^{[\mu]}(t)
V^{[\mu]\dag}_{jb}(t).
\ee
Then by using the singular value $\sigma$ and the unitary matrices $U$ and $V$, tensors are created as
\be
T_{ijkl}(t)
\equiv
\sqrt{
\sigma_i^{[0]}(t)
\sigma_j^{[1]}(t)
\sigma_k^{[0]}(t-1)
\sigma_l^{[1]}(t)
}
\sum_{a=1}^{N}
U_{ai}^{[0]}(t)
U_{aj}^{[1]}(t)
V_{ak}^{[0]\ast}(t-1)
V_{al}^{[1]\ast}(t)
\left.\frac{d\tilde\phi(\xi)}{d\xi}\right|_{\xi=\xi_a}.
\ee
The bond dimensions for the tensor indices $i,j,k,l$ are determined adaptively
according to a required precision,
e.g., for a parameter set ($\lambda=10$ and $\gamma=3$) in the next section,
when a required relative precision is $10^{-4}$,
the maximum bond dimension is around $200$.
In terms of the tensors, the path integral is now given by
\be
{\cal Z}
\approx
\sum_{\{i,j,k,l\}}
\prod_{t,n}T_{ijkl}(t).
\label{eqn:Z_tensor_network}
\ee
The resulting tensor network is sketched in Fig.~\ref{fig:tensor_network} which has a layer structure.
As seen in Fig.~\ref{fig:SVD},
the correlation is strong in the Minkowski region while it is weak in the Euclidean case.

\section{Numerical results}
With proper field insertions, the time-correlator is defined by
\be
\langle
\phi_{t,n}\phi_{0,0}
\rangle
\equiv
\frac{
\int
[d\phi]
\phi_{t,n}\phi_{0,0}
e^{iS_{\rm lat}[\phi]}}
{
\int
[d\phi]
e^{iS_{\rm lat}[\phi]}
}.
\ee
The numerator can also be represented by a tensor network with a couple of impurity tensors.

\begin{figure}[t!]
\begin{center}
\begin{tabular}{c}
\includegraphics[width=10cm,pagebox=cropbox,clip]{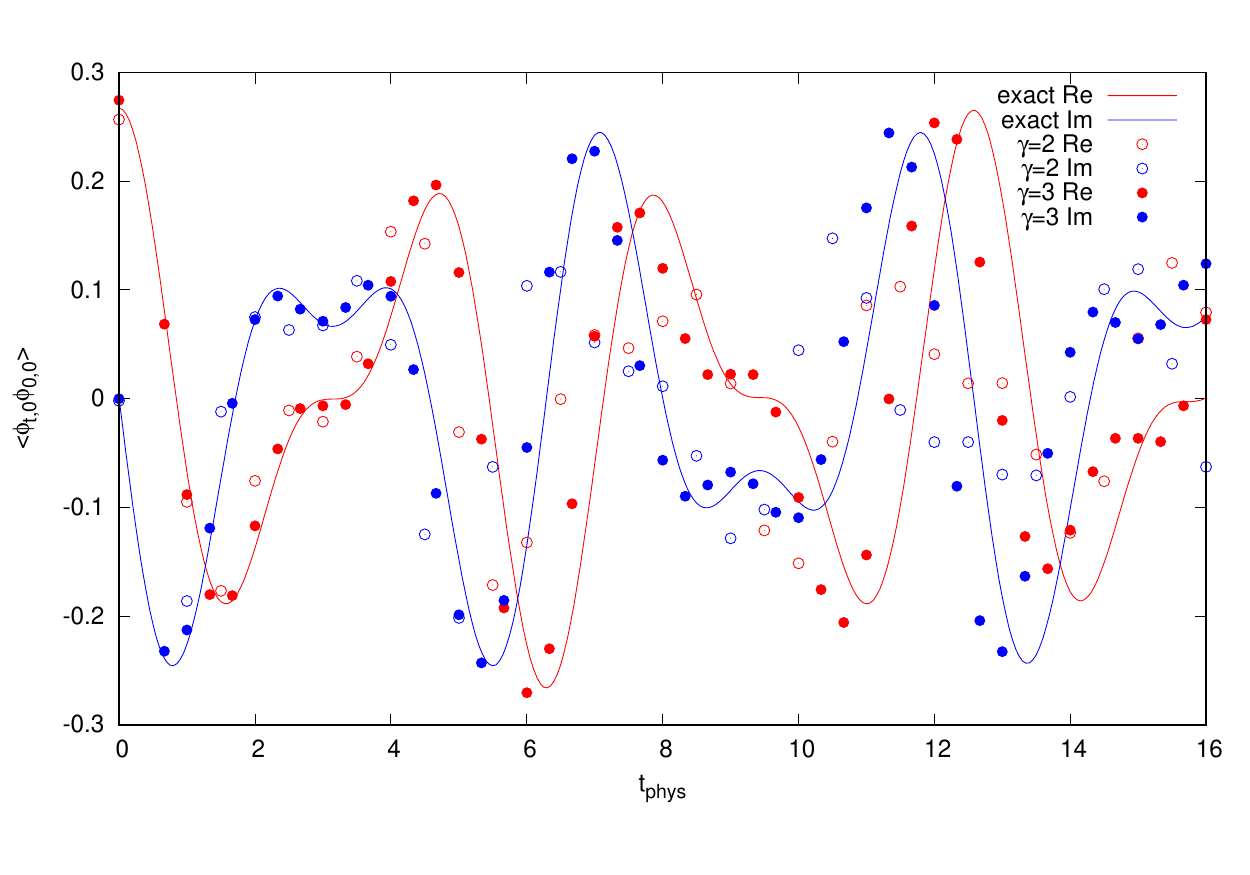}
\end{tabular}
\end{center}
\vspace{-12mm}
\caption{
The time-correlator $\langle
\phi_{t,0}\phi_{0,0}
\rangle
$ for $N_{\rm S}=2$ as a function of $t_{\rm phys}=t/\gamma$.
The parameters are $m=1$, $\lambda=10$, $\beta=4$, $T=16$ and $\gamma=2,3$.
The real part and the imaginary part of the correlator are represented by the red and blue points respectively.
The solid curve is the exact result obtained by the exact diagonalization.
}
\label{fig:correlator_Ns2}
\end{figure}

In Fig.~\ref{fig:correlator_Ns2}, we plot the numerical results of the time-correlator for $N_{\rm S}=2$.
Since the system is so small, a coarse-graining is not applied and the contraction (cf. the RHS of eq.(\ref{eqn:Z_tensor_network})) is done exactly.
Our result agrees with the exact one in the early stage while their difference gradually incearses as time evolves.
The reason for the deviation is considered as discretized time effects,
in fact, the deviation is smaller for larger
$\gamma$.

\begin{figure}[t!]
\begin{center}
\begin{tabular}{c}
\includegraphics[width=8.cm,pagebox=cropbox,clip]{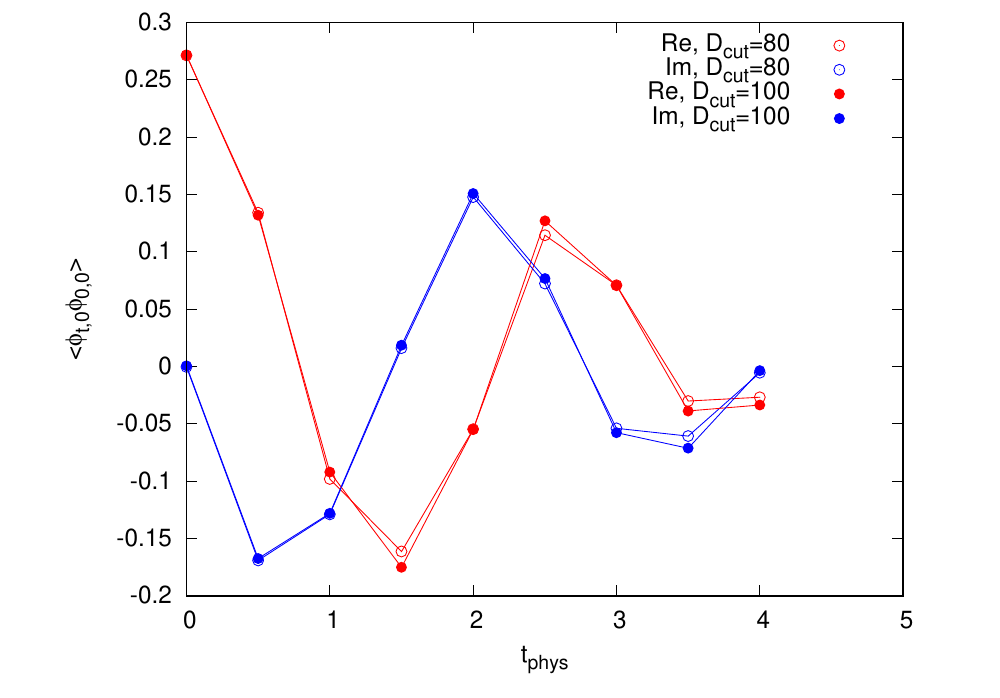}
\end{tabular}
\end{center}
\vspace{-5mm}
\caption{
The time-correlator  $\langle
\phi_{t,0}\phi_{0,0}
\rangle
$ for $N_{\rm S}=64$ as a function of $t_{\rm phys}=t/\gamma$.
The parameters are $m=1$, $\lambda=16$, $\beta=8$, $T=4$ and $\gamma=2$.
$D_{\rm cut}$ is a truncation order for a corse-graining scheme.
}
\label{fig:correlator_Ns64}
\end{figure}

Figure \ref{fig:correlator_Ns64} shows the correlator at $N_{\rm S}=64$.
To attain the large volume, we employ a coarse-graining scheme whose
truncation order is denoted by $D_{\rm cut}$ and larger value of $D_{\rm cut}$ is expected to give a better precision.
Here, two cases $D_{\rm cut}=80$ and $100$ are plotted to see a stability against the truncation process.
At the scale, the result is not so sensitive to $D_{\rm cut}$,
thus 
this indicates
that we can reliably compute the correlator with the parameter set shown here.

\section{Summary and future directions}

We have derived a tensor network representation for
the real scalar field theory in the closed-time formalism and
observed that the resulting network has a non-uniform entanglement (correlation) structure.
In this paper, we exclusively consider 1+1 dimensional system, but an extension to higher dimensional system is straightforward.
By using the tensor network representation, we numerically evaluate the time-correlator.
We confirm that the correlator is correctly calculated
by comparing with the exact result at the small spatial size $N_{\rm S}=2$.
As a demonstration, we also show the correlator at relatively larger volume $N_{\rm S}=64$
and confirm that the result is stable against the change of the truncation order of the coarse-graining.

As observed, 
the entanglement tends to be stronger for weak coupling and near the time continuum limit.
Before starting serious calculations, thus, one should find a better way of making an initial tensor network with small truncation errors
even for such parameter regions.
Furthermore, finding a better coarse-graining scheme is also an important task for future applications.
In any case, we now have a non-perturbative computation tool to evaluate the time-correlator
thus we may try to directly compute the spectral function and the distribution function near future.


\vspace{3mm}
This work is supported in part by
JSPS KAKENHI Grant Numbers JP17K05411
and
JP21K03531.

\end{document}